\documentclass[aps,twocolumn]{revtex4}

\pdfoutput=1

\usepackage{eurosym}
\usepackage{amsfonts}
\usepackage{amsmath}
\usepackage{amssymb,epsf}
\usepackage{color}
\usepackage{graphicx}
\usepackage{epstopdf}
\usepackage{float}
\usepackage{caption}
\usepackage{subfig}
\usepackage{hyperref}
\usepackage{mathtools}
\usepackage{lipsum}

\makeatletter
\newbox\sf@box
{\def\sf@one{#1}%
	\def\sf@two{#2}%
	\setbox\sf@box\hbox
	\bgroup}%
{ \egroup
	\ifx\@empty\sf@two\@empty\relax
	\def\sf@two{\@empty}
	\fi
	\ifx\@empty\sf@one\@empty\relax
	\subfloat[\sf@two]{\box\sf@box}%
	\else
	\subfloat[\sf@one][\sf@two]{\box\sf@box}%
	\fi}
\makeatother

\makeatletter
\renewcommand{\p@subfigure}{\thefigure-}
\makeatother

\definecolor{aogreen}{rgb}{0.0, 0.5, 0.0}

\def\ketm#1{  \left\vert  #1   \right\rangle   }

\def\bram#1{  \left\langle  #1   \right\vert   }

\begin{document}

\title{Increasing the efficiency of quantum walk with entangled qubits}
\author{S. Panahiyan$^{1,2,3}$ \footnote{%
		email address: shahram.panahiyan@uni-jena.de} and S. Fritzsche$^{1,2,3}$%
	\footnote{%
		email address: s.fritzsche@gsi.de} }

\affiliation{$^1$Helmholtz-Institut Jena, Fr\"{o}belstieg 3, D-07743 Jena, Germany  \\
	$^2$GSI Helmholtzzentrum f\"{u}r Schwerionenforschung, D-64291 Darmstadt, Germany \\
	$^3$Theoretisch-Physikalisches Institut, Friedrich-Schiller-University Jena, D-07743 Jena, Germany}

\begin{abstract}
We investigate how arbitrary number of entangled qubits affects properties of quantum walk. We consider variance, positions with non-zero probability density and entropy as criteria to determine the optimal number of entangled qubits in quantum walk. We show that for a single walker in one-dimensional position space, walk with three entangled qubits show better efficiency in considered criteria comparing to the walks with other number of entangled qubits. We also confirm that increment in number of the entangled qubits results into significant drop in variance of probability density distribution of the walker, change from ballistic to diffusive (suppression of quantum propagation), localization over specific step-dependent regions (characteristic of a dynamical Anderson localization) and reduction in entropy on level of reaching the classical walk's entropy or even smaller (attain deterministic behavior). In fact, we see that for large number of the entangled qubits, quantum walk loses most of its properties that are celebrated for but still show characterizations which are genuinely diffident comparing to classical walk.    
\end{abstract}

\maketitle

Quantum walks (QWs) are universal computational primitives \cite{Lovett,Childs}. They are used to develop quantum algorithms \cite{Ambainis,Shenvi}, simulate quantum systems \cite{Mohseni}, engineer/prepare quantum states \cite{Innocenti}, quantum  machine learning \cite{Paparo} and design quantum neural networks \cite{Schuld1,Schuld2}. The QWs have been realized experimentally with Bose-Einstein condensate \cite{Dadras}, optical-network \cite{Barkhofen}, photons \cite{Schreiber,Bian}, ions \cite{Zahringer} and atoms \cite{Karski}. 

There are several approaches to generalize a one-dimensional QW with single walker and two integral degrees of freedom. One way is to focus on high dimensional position and coin spaces for the walk. Although the high dimensional position space has its own advantages, it has several considerable drawbacks. For example, one has to change the architecture of experimental setup on high level to integrate high dimensional position space into QWs. Therefore, we keep dimension of position as one and pursue high dimensional coin space. 

One can obtain high dimensional coin space through two schemes: whether to have a single qubit with many states or a system of two-stated qubits which are entangled with each other. In our study, we are interested in latter one. This is because increasing the dimension of the entangled states of a system can make its non-classical correlations more robust to the presence of noise and other detrimental effects \cite{Collins}. In addition, the high dimensional entangled state implies a greater potential for applications in quantum information processing \cite{Vertesi,Martin} and communication \cite{Cerf}. 

So far, there have been several studies where the walks with two entangled qubits are addressed \cite{Venegas,Liu,Liu2012,Panahiyan}. A general investigation for arbitrary number of entangled qubits yet to be made. In this paper, we consider a single walker in one dimensional position space with $2^{n}$ internal degrees of freedom. The internal degrees of freedom is a result of $n$ qubits that are entangled with each other. The main goal here is to study the properties of the QWs for $n$ entangled qubits and find the optimal number of entangled qubits (NEQs) to increase the walk's efficiency. To determine the optimum number of entangled qubits for the walker, we investigate the properties of the walker in position space where we have the possibility to make comparison between walks with different NEQs as well as classical walk (CW). These properties include symmetry of the distribution, its variance, the number of positions occupied by walker's wave function in position space and entropy associated to position space. 

Formally, the evolution of the discrete QWs is determined by a coin, $\widehat{C}$, and a shift, $\widehat{S}$, operator. In each step, the coin operator modifies the amplitudes of walker's wave function in coin space and creates a superposition of internal states. On the other hand, the shift operator moves the walker in position space based on its internal state. We consider the coin operator to be 

\begin{eqnarray}
\widehat{C} & = & \widehat{C'}^{\otimes n},
\end{eqnarray}
where $n$ is the NEQs, $\otimes$ is tensor product and $\widehat{C'}$ is Hadamard operator given by

\begin{eqnarray*}
\widehat{C'} & = & \frac{1}{\sqrt{2}}( \ketm{0}_{C} \bram{0} \:+\: \ketm{0}_{C} \bram{1} 
+\: \ketm{1}_{C} \bram{0} \:-\:  \ketm{1}_{C} \bram{1}).                    
\end{eqnarray*}
in which, we have $\ketm{0}_{C} \equiv \begin{bmatrix} 1 \\ 0 \end{bmatrix}_{C}$ and $\ketm{1}_{C} \equiv \begin{bmatrix} 0 \\ 1 \end{bmatrix}_{C}$.

The dimension of coin space is determined by the NEQs which is $2^{n}$. Therefore, for $n$ entangled qubits, $\widehat{C}$ is a $2^{n} \times 2^{n}$ matrix. It is notable that the application of the considered coin on walker's internal state affect each qubit identically in every step. We consider the shift operator to be in the form of 

\begin{eqnarray}
\widehat{S} & = & \left\{
\begin{array}{c} 
\ketm{0}_{C}^{\otimes n} \otimes \ketm{x}_{P} \Longrightarrow \ketm{0}_{C}^{\otimes n} \otimes \ketm{x+1}_{P} \\[0.4cm]

\ketm{1}_{C}^{\otimes n} \otimes \ketm{x}_{P} \Longrightarrow \ketm{1}_{C}^{\otimes n} \otimes \ketm{x-1}_{P} \\[0.4cm]
\mathrm{Otherwise} \otimes \ketm{x}_{P} \Longrightarrow  \mathrm{Otherwise} \otimes \ketm{x}_{P}
\end{array}  
\right.  .                   \label{shift}
\end{eqnarray}
where $\ketm{x}_{P}$ is the position of walker. The Hilbert space of position is spanned by $\{ \ketm{x}_{P}: x\in \mathbb{Z}\}$. This shift operator provides three possibilities for walker's movement in position space except in case of $n=1$. For single qubit, the third possibility in Eq. \eqref{shift} is removed and the walker moves only one unit to the right and/or left. The walk is realized by successive application of the coin-shift operator on an initial state of walker for arbitrary times, $T$, 

\begin{eqnarray}
\ketm{\psi}_{Final} & = &   [\widehat{S} \widehat{C}]^{\:T}\: \ketm{\psi}_{Initial},
\end{eqnarray} 
in which  

\begin{eqnarray}
\ketm{\psi}_{Initial} & = & \frac{1}{\sqrt{2}} (\ketm{0}_{C}^{\otimes n} \:+\:  \ketm{1}_{C}^{\otimes n})\otimes \ketm{0}_{P},     \label{int}                                                          
\end{eqnarray}
where we have localized the walker in the origin of position space. For $n=2$, we have a Bell state for coin space while for $n>2$, classes of GHZ states are resulted. One can experimentally realize this walk by combining two different experiments where in one; QWs are implemented by cavities in which internal states are given by two-stated ions \cite{Sanders,Flurin} and in the other one; entangled qubits (ions) and high controllability over their states are included in cavities \cite{Casabone}. 

For the walk considered in this paper, we have two types of entanglement: I) position and coin spaces are entangled with each other. II) qubits that build up coin space, i. e. internal degrees of freedom of the walk(er), are entangled with each other. Previously, it was pointed out that distinct properties of the QWs, comparing to CW, are originated in entanglement between the position and coin spaces. In fact, to obtain classical like behavior in QWs, one can make weak measurements in coin space of walk at each step \cite{Brun}. This results into elimination of entanglement between position and coin spaces. The entanglement between qubits has not been explored in details for QWs. Here, we concentrate on this entanglement. 

We have done simulation of walk for up to seven entangled qubits (Figs. \ref{Q1} - \ref{Q7}) while in the appendix,  we have given the analytical description for arbitrary NEQs. We also present the simulation for CW (Fig. \ref{CW}) to have a more comprehensive framework for making comparison and point out to a few characterizations if the coin space arises from a large NEQs.

The variance and/or standard deviation is our first factor/property to determine optimal NEQs. The variance/standard deviation is one of properties that distinguishes QW from its classical counterpart. In fact, it was due to this property that speedup in QW algorithms, comparing to CW ones, was concluded. Here, we see that the variance decreases with NEQs (see Fig. \ref{Fig1}). This indicates that increment in NEQs results into destructive contributions on walker’s most left and right hand side probability densities. The standard deviation is the square root of variance. Therefore, one can conclude that standard deviation of the walk with single qubit is larger than the other ones and also CW. In general, we have the following relation for the standard deviation 

\begin{eqnarray*}
\sigma_{CW} < \sigma_{\mathrm{7-qubits}}  <  \sigma_{\mathrm{6-qubits}}  <  \sigma_{\mathrm{5-qubits}}   <  
\\
\sigma_{\mathrm{4-qubits}}  <  \sigma_{\mathrm{3-qubits}}  <   \sigma_{\mathrm{2-qubits}}  < \sigma_{\mathrm{1-qubit}},                                               
\end{eqnarray*}
which shows that increment in NEQs results into slower spread of the walker's wave function in position space. 

Previously, it was pointed out that walk with a single qubit and Hadamard coin is ballistic where its speed is almost half of a free particle with unit velocity \cite{Portugal}. The large NEQs results into walker losing its ballistic nature and becomes more diffusive similar to CW. But here, there is a significant difference between CW and walks with large NEQs; although the standard deviations of walks with large NEQs come close to CW, their expected value of position differs completely. For CW, expected value of position is zero \cite{Portugal}. In contrast, for walks with large NEQs, expected value of position is within right and/or left hand sides of the origin and it changes from one step to another one.   

Next optimal factor is symmetrical/asymmetrical evolution of the walker's probability density distribution. This property determines the contributions of different protocols involved in QW. Careful examination of the probability density distributions shows that its symmetry depends on NEQs (see Fig. \ref{Fig1}). Walks with even NEQs have symmetrical probability density distribution while we observe the asymmetrical distributions for odd NEQs. 

Previously, in case of single qubit walk, it was argued that the source of asymmetrical distribution is only rooted in the considered initial state (independent of coin operator). If in the initial state, the internal states and their evolutions through the walk are isolated from one another, probability density distribution of walk becomes symmetrical \cite{Portugal}. Here, we see that for odd NEQs, this argument is valid whereas for even entangled qubits, this is not the case. The reason is that the symmetry of distribution is determined not only by initial state, but also the coin-shift operator (contrary to present arguments in the literature). One can express the evolution of the walk in terms of eigenvalues and eigenvectors of shift-coin operator and initial state (please see appendix). The eigenvalues and eigenvectors of shift-coin operator have specific factors which are different for odd and even NEQs. These factors determines whether for having symmetrical distribution, the evolution of internal states of the walker in initial state needs to be isolated or not. We see that such isolation is necessary for odd NEQs.

\begin{figure*}[!htbp]
	\centering
	{\begin{tabular}{cccc}		
	\subfloat[One qubit]{\label{Q1}\includegraphics[width=0.22\linewidth]{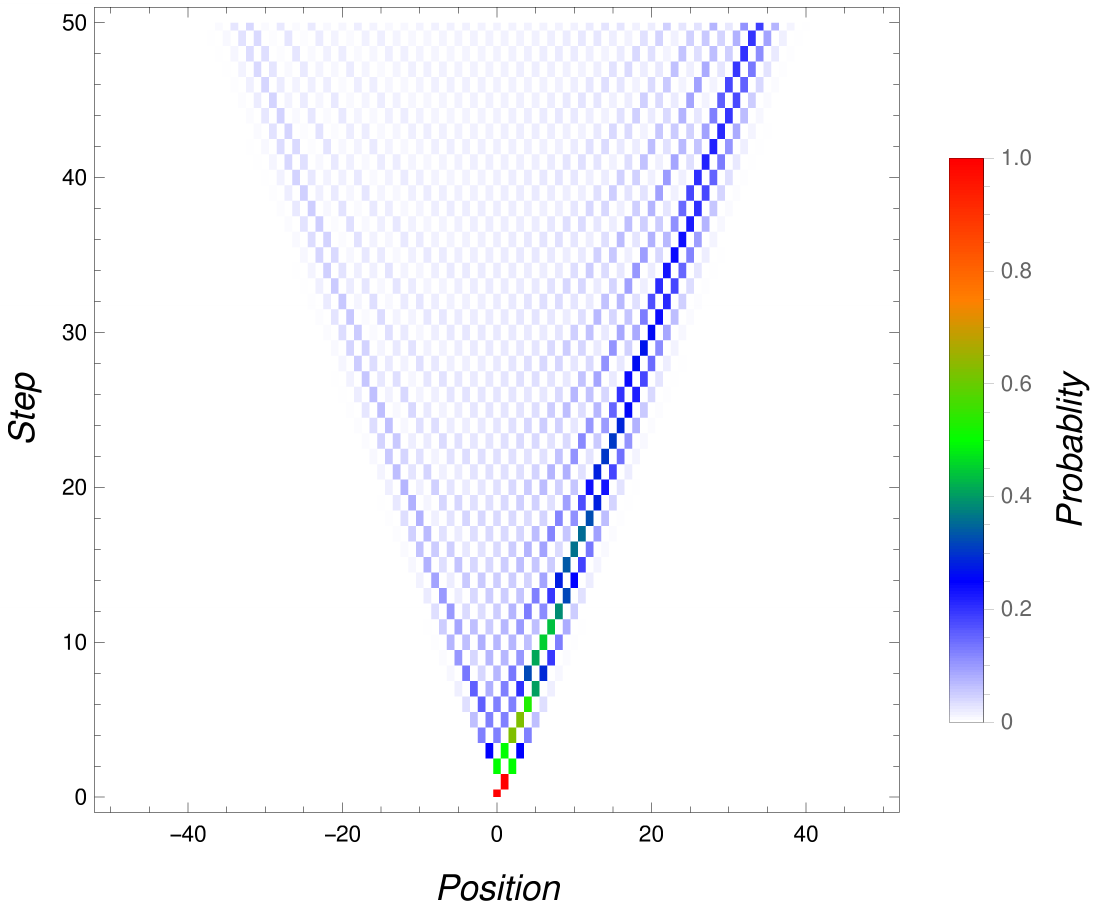}}\quad 
	\subfloat[Two entangled qubits]{\label{Q2} \includegraphics[width=0.22\linewidth]{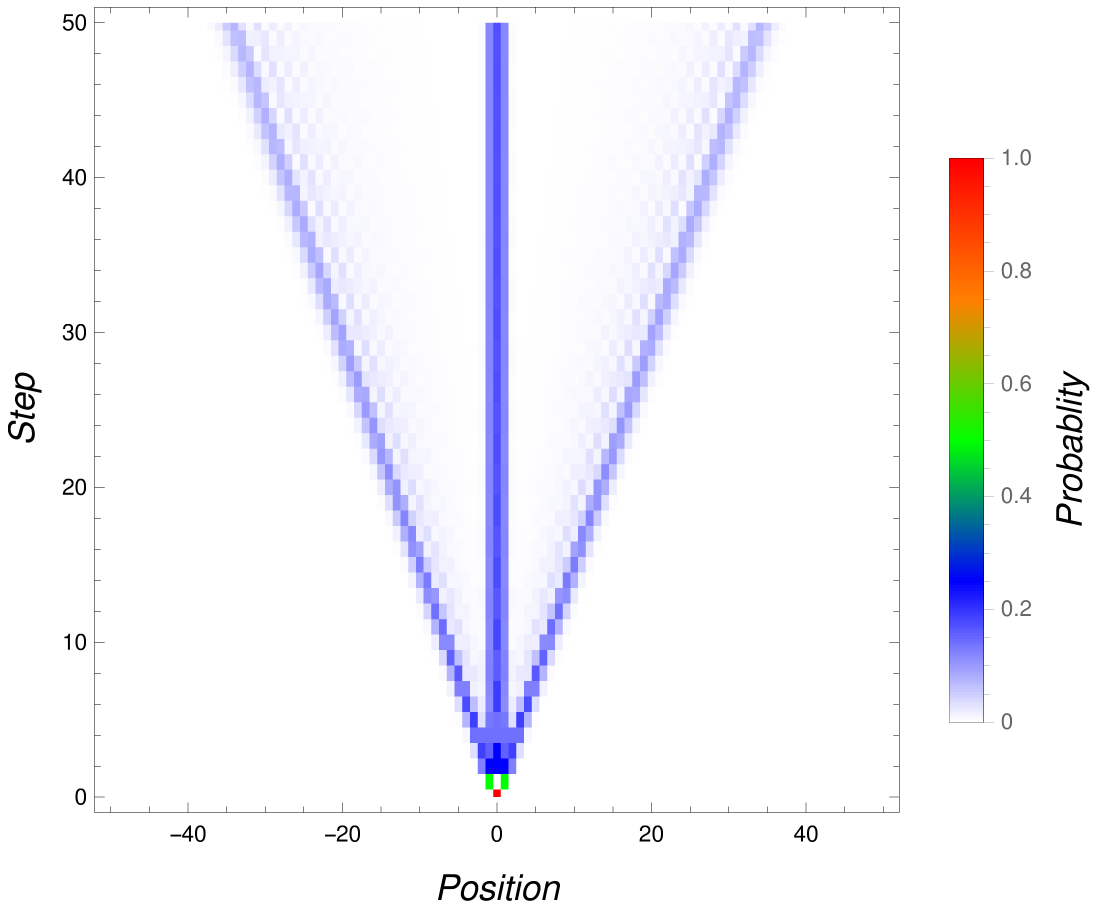}}\quad 
	\subfloat[Three entangled qubits]{\label{Q3} \includegraphics[width=0.22\linewidth]{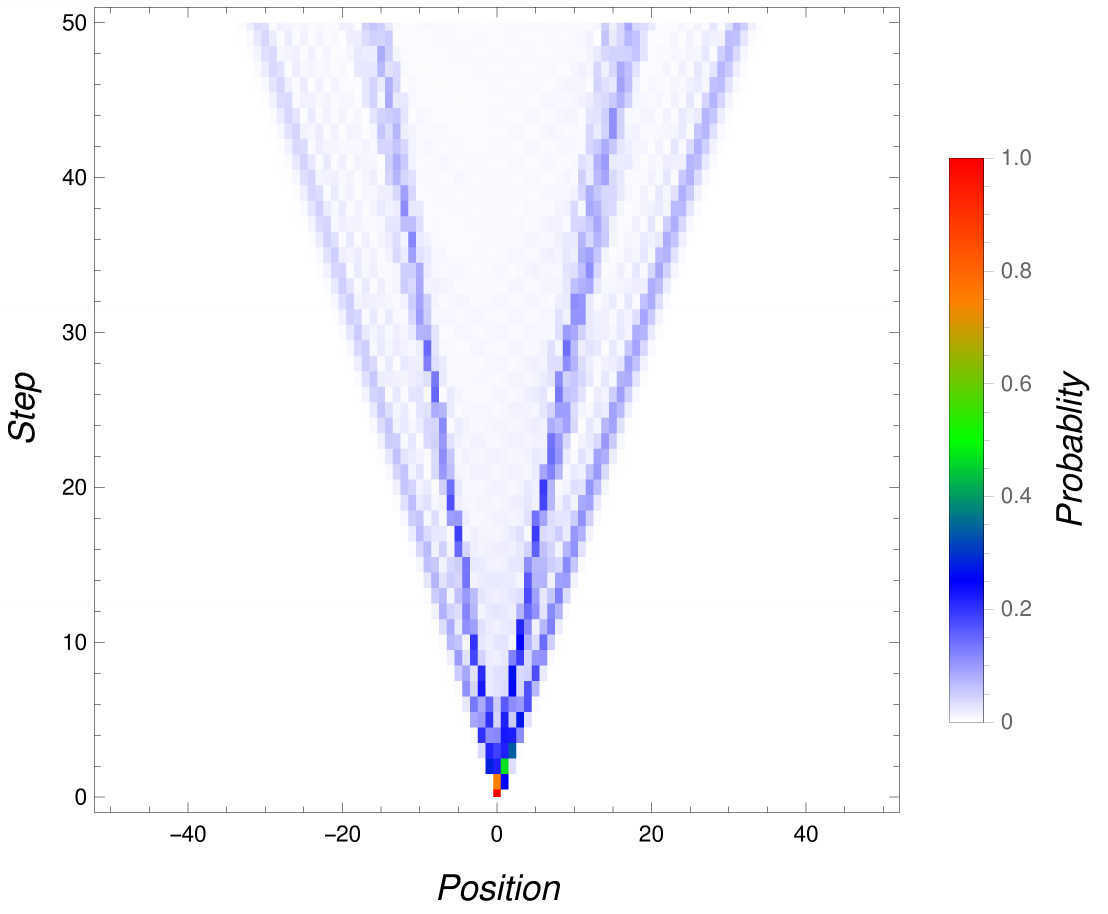}}\quad 	   		
	\subfloat[Four entangled qubits]{\label{Q4} \includegraphics[width=0.22\linewidth]{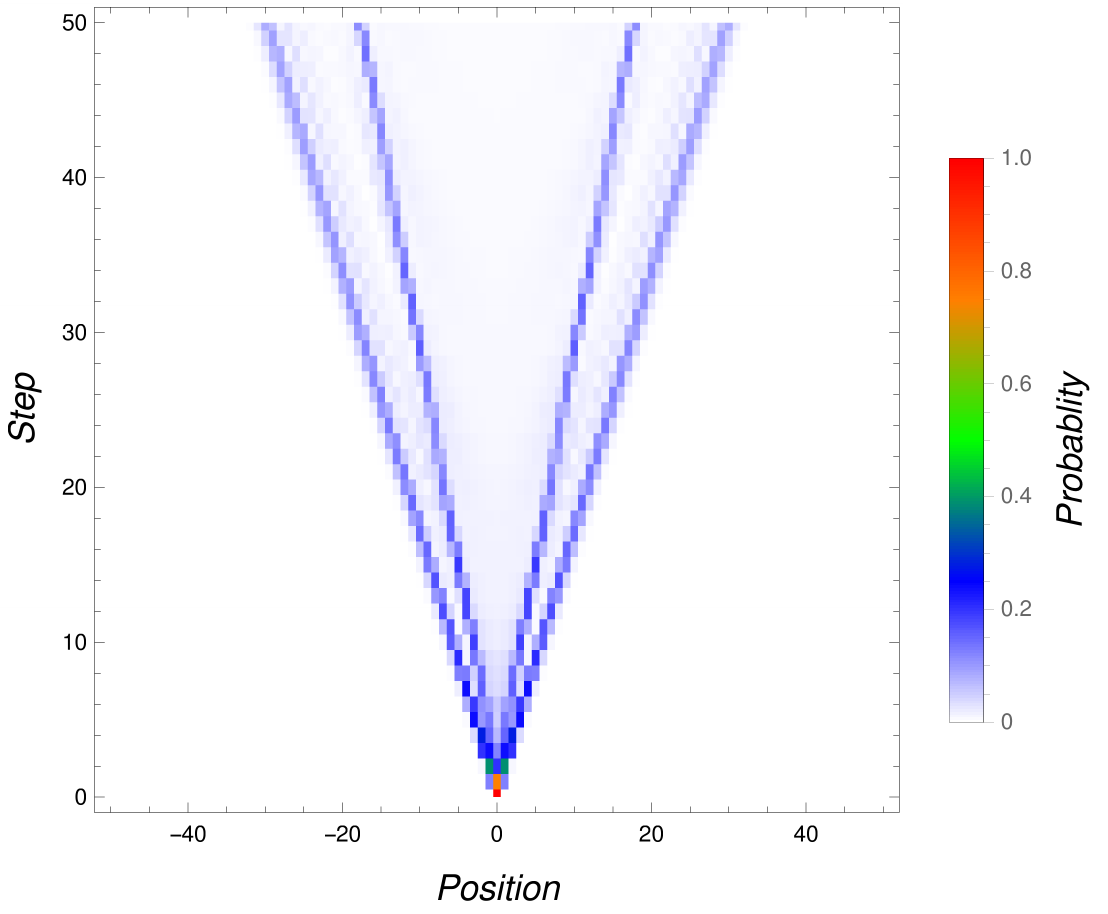}}\\ 
	\subfloat[Five entangled qubits]{\label{Q5} \includegraphics[width=0.22\linewidth]{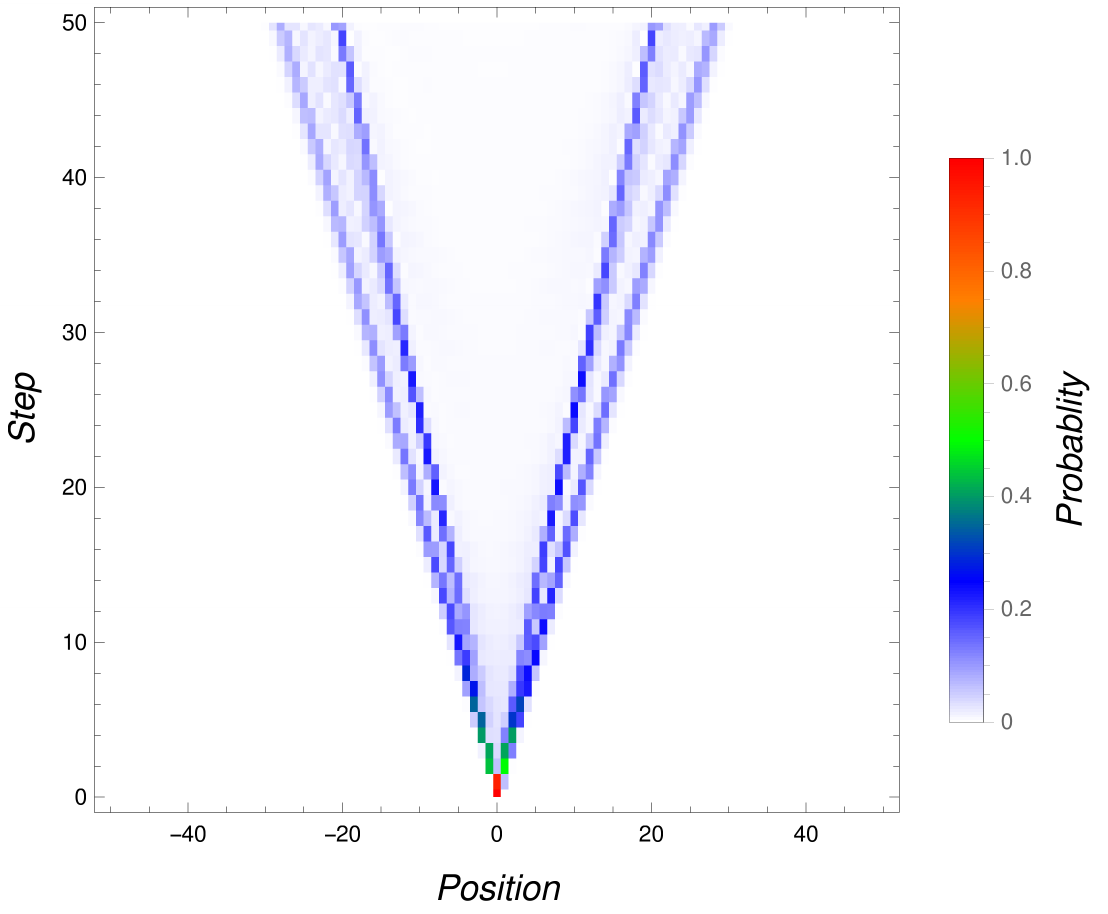}}\quad  
	\subfloat[Six entangled qubits]{\label{Q6} \includegraphics[width=0.22\linewidth]{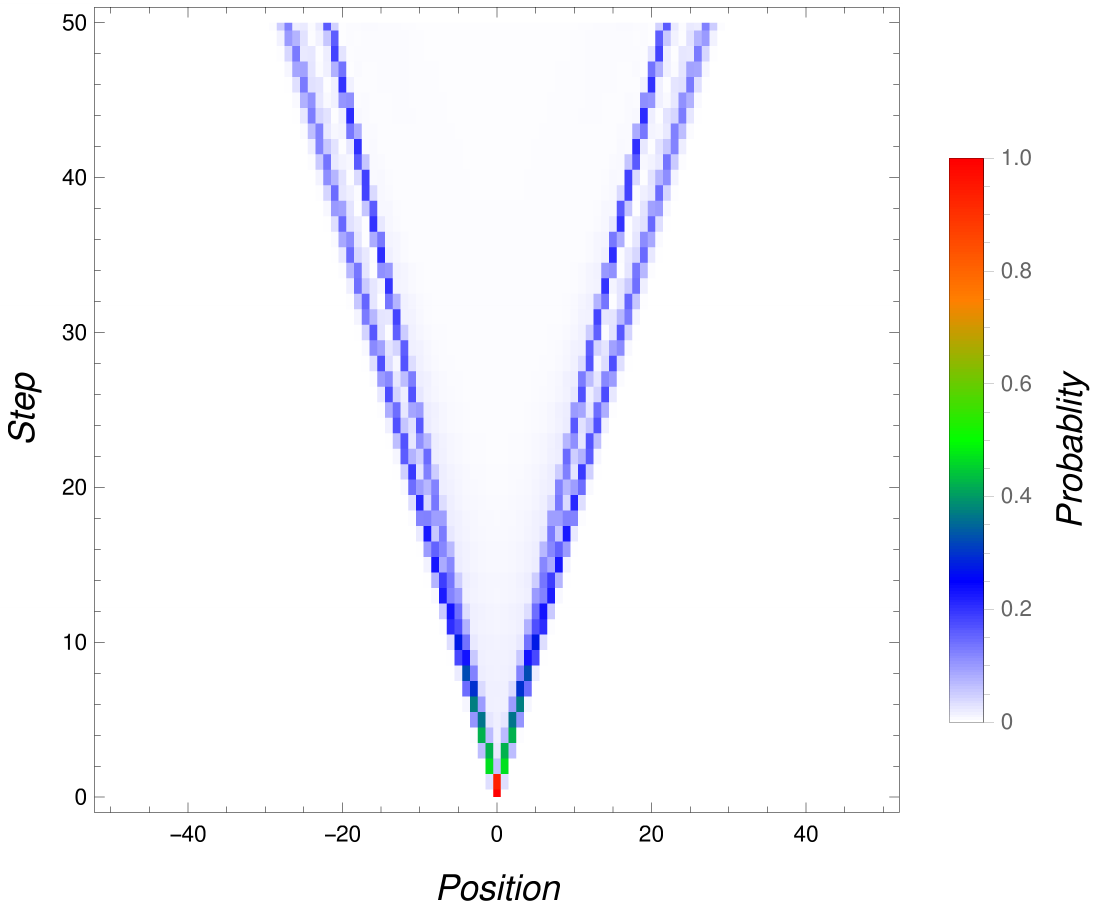}}\quad 
	\subfloat[Seven entangled qubits]{\label{Q7} \includegraphics[width=0.22\linewidth]{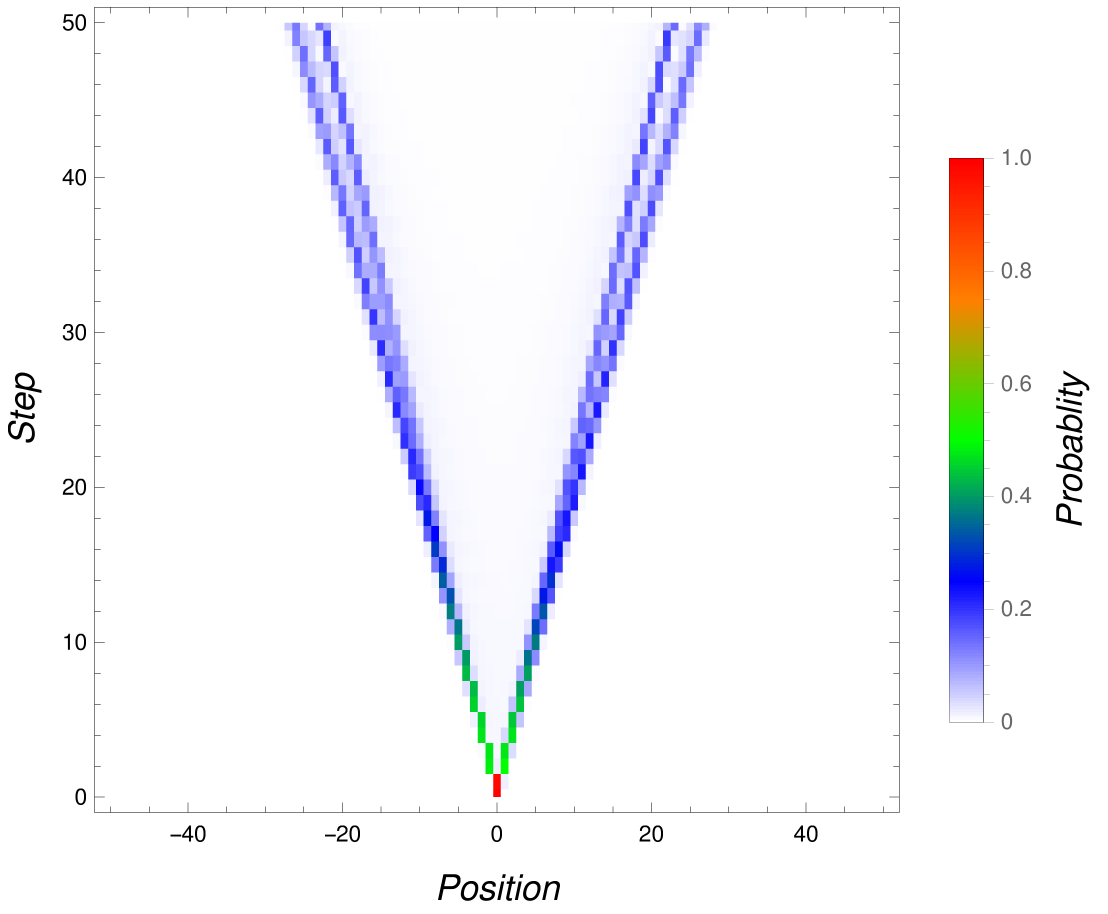}}\quad 
	\subfloat[Classical walk]{\label{CW} \includegraphics[width=0.25\linewidth]{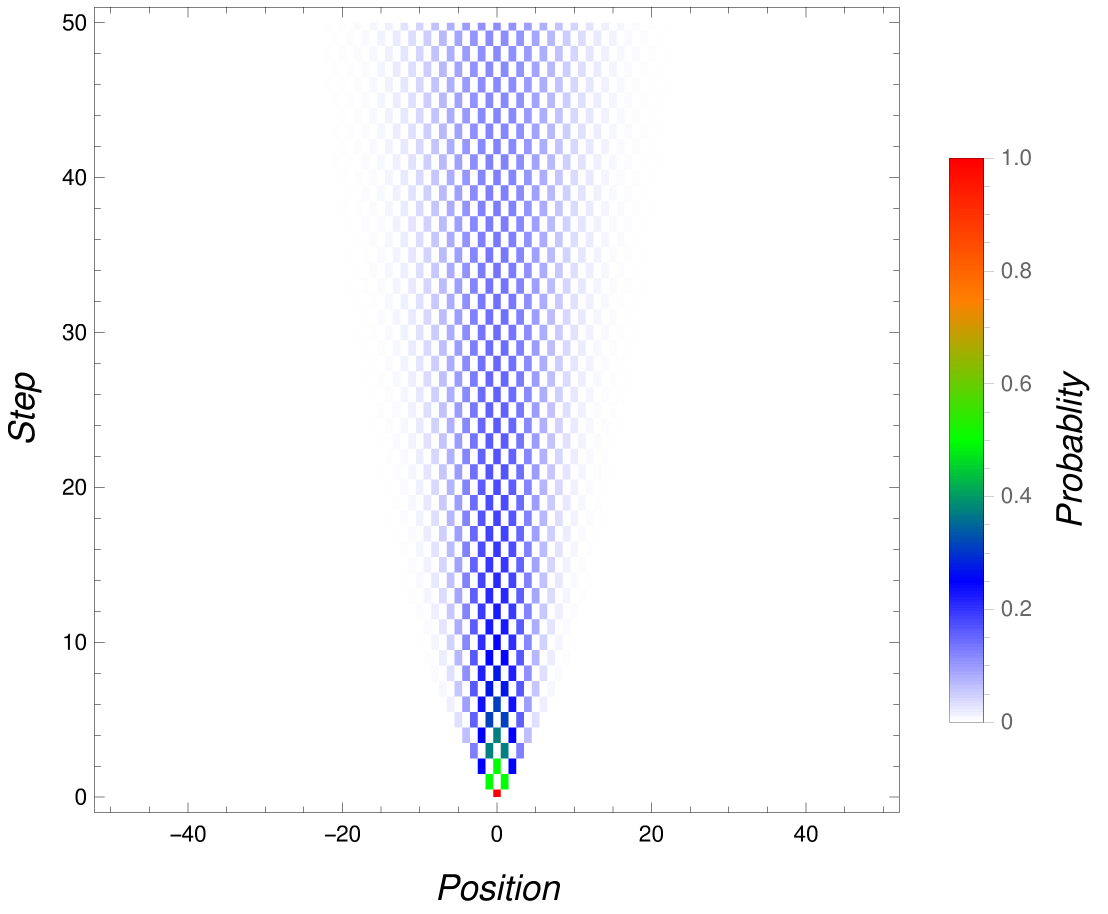}}
	\end{tabular}}				
	\caption{Probability density distribution in position space for $50$ subsequent steps.} \label{Fig1}
\end{figure*}	
\begin{figure*}[htb]
\centering
    {\begin{tabular}[b]{cc}%
    \subfloat[\label{NP} Number of positions with probability densities bigger than $10^{-4}$.]{\includegraphics[width=0.39\linewidth]{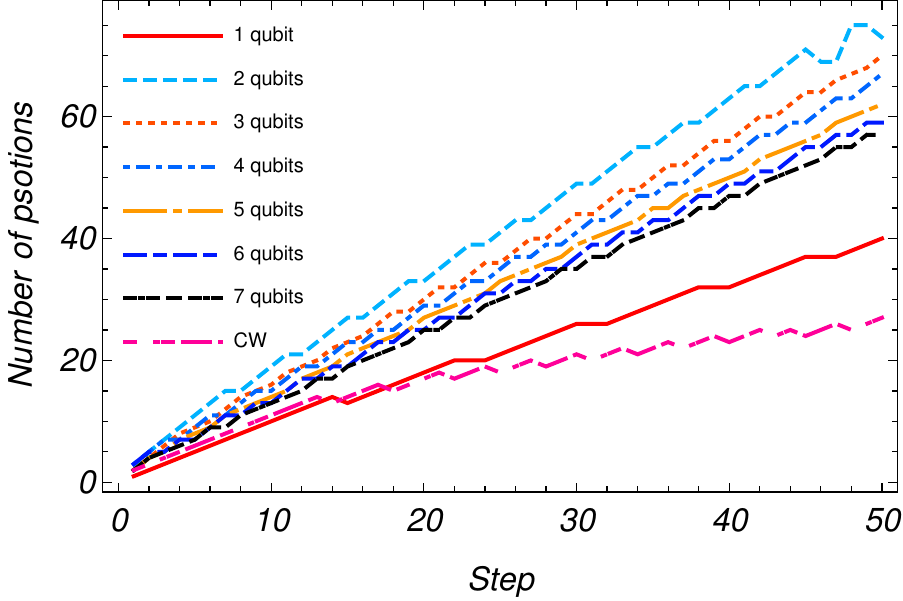}}\
	\subfloat[\label{EN} Total entropy of walk as a function of steps.]{\includegraphics[width=0.39\linewidth]{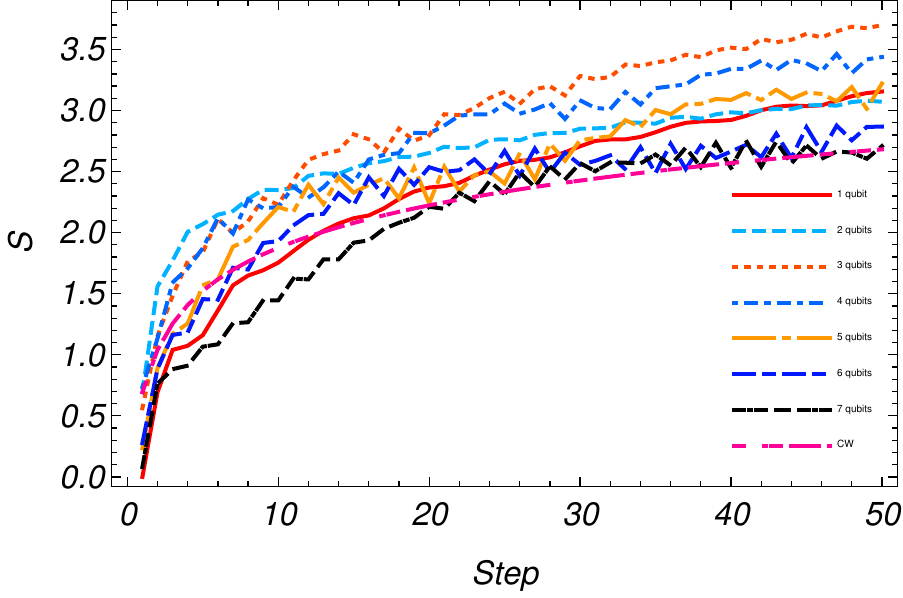}}
	\end{tabular}}				
	\caption{Number of positions with non-zero probability density and entropy of walk for $50$ subsequent steps.} \label{Fig2}
\end{figure*}	

The next factor for determining the optimal NEQs is number of positions with non-zero probability density. Here, we make a distinction between variance and this factor due to applications of QWs in developing quantum algorithms and designing quantum neural networks. In designing algorithms such as search algorithms, the accessibility to blocks where search is done, is one of the crucial factors. In the walk with a single qubit, larger variance (faster speed for walker's wave function propagation) in position space is observed. But this is at the expense of neglecting half of the positions available for walker in each step \ref{NP}. In contrast, for the walks with $1<n<n'$, we see the smaller variance but significantly larger number of positions with non-zero probability density \ref{NP}. Effectively, walk with two entangled qubits has the largest number of positions with non-zero probability density. The number of positions with non-zero probability density is a decreasing function of the NEQs and for $n'<n$ (sufficiently large $n$), the walker's number of positions with non-zero probability density could become equal or smaller than the one for walk with single qubit. This is due to fact that increment in NEQs results into a more localized probability density distribution over smaller regions of positions. The smallest number of positions with non-zero probability density belongs to CW. It is expected that by increasing $n$ in QWs, its number of positions with non-zero probability density become equal with CW one's. While such equality could be obtained, one should remember that for QWs, the positions with non-zero probability density changes from one step to another one and these positions drift away from origin whereas, for CW, the positions with non-zero probability density are highly concentrated around the origin.

Entropy is our final factor to determine the optimal NEQs. The entropy measures the uncertainty in state of a system. For QW, large entropy in position space indicates that walker's probability density distribution in position space is more homogeneous (closer to uniform walk with maximum entropy). Therefore, the walk is less deterministic and it is harder to guess the walker's position after $T$ step. In contrast, smaller entropy show a more deterministic nature for walker's behavior in position space where it is more possible to guess walker's position at each step. The minimum entropy is zero where probability density of a single position becomes unit. Therefore, one can say where exactly the walker is located. Previously, it was shown that a deterministic behavior for walker in position space is a destructive consequence for QWs applications in quantum algorithm and communication \cite{Keating}. Therefore, the large entropy shows better efficiency for QWs applications in quantum algorithm and communication. 

In QWs, since the Hilbert space of the walk is made out of two subspaces of position and coin, there are two definable entropies of coin (known as entanglement entropy) and position spaces. Both of these entropies depend on dimension of coin and position spaces. Therefore, only the entropy of position space provides a comparison tool between the walks with different NEQs and CW as well. It should be noted that entanglement entropy is an unique property of QWs which is absent in CW. 

In multipartite systems, one can use the method of reduced density matrix to calculate entropy associated to different parts of the system. At step $T$ of the walk, the density operator is $\hat{\rho}_{T} = \ketm{\psi}_{T} \prescript{}{T}{\bram{\psi}}$. The $ \ketm{\psi}_{T}$ is given by two subspaces of coin and position spaces. To find the entropy associated to one of these subspaces, one should take the partial trace over the other space. For example to obtain entropy in position space, the partial should be taken over coin space, $\hat{\rho}^{P}_{T}=\text{Tr}_{C}(\hat{\rho}_{T})$. It is a matter of calculation to find the entropy of position space at time $T$ as

\begin{eqnarray}
S^{P}_{T} &=& -\text{Tr}(\hat{\rho}^{P}_{T} Log\hat{\rho}^{P}_{T})=-\sum_{n}P_{n}LogP_{n},  \label{von}
\end{eqnarray}
where $P_{n}$ are eigenvalues of Hermitian matrix with the element $\hat{\rho}^{P}_{T}$ and $n$ is the position. Here, we put our focus on the largest step given in Fig. \ref{EN} but the arguments are generally valid for long term limit. 

The highest value of the entropy belongs to walk with three entangled qubits \ref{EN}. For cases with more than three entangled qubits, the entropy is a decreasing function of the NEQs. We observe that for seven entangled qubits, the entropies of the quantum and classical walks become equal. It is expected that for larger NEQs, the entropy would even decrease more than CW. Here, we have two exceptions: walk with single and two entangled qubits. Although entropy of single qubit is small at early stage of walk, it has a large growth rate. This is because walk with single qubit has the fastest spread of the walker's wave function in position space. For two entangled qubits, the entropy is intermediate. The reason is that while the walk with two entangled qubits has relatively good speed for propagation of the walker's wave function in position space, the major probability density is observed at starting position of the walk. 

The sharp drop in entropy as NEQs increases shows that for large NEQs, walker's behavior becomes more deterministic. For sufficiently large NEQs, the entropy would converge to zero which indicates that walker shows characteristic of Anderson localization \cite{Keating}. But, the regions of this localization is step-dependent which makes it a dynamical Anderson localization. Such behavior is rooted in the fact that for large NEQs, both variance and number of positions with non-zero probability density decrease significantly (suppression of quantum propagation in position space).  It is worthwhile to mention that while the entropy in position space drops in case of increasing the NEQs, the entropy in coin space increases. This is because in each application of the coin on the walker, the amplitudes in coin space become more distributed among the internal states creating a superposition of them all. Since the number of internal states is given by $2^n$, the entropy of coin space becomes an increasing function of the NEQs. 

In our study, we found that walks with three entangled qubits has relatively a very good variance, one of the highest number of positions with non-zero probability density and the highest entropy in position space. These factors indicate that for algorithm and quantum communication purposes, walks with three entnagled qubits are the most efficient ones. The single and two entangled qubit walks also show good performance but, the single qubit case has a very small number of positions with non-zero probability density at each step and walks with two entangled qubits show more deterministic behavior.  

In conclusion, we showed that for walks with large number of entangled qubits, the quantum interference due to entanglement between qubits shapes the walker's probability density distribution into a more localized one over two regions and suppresses the quantum propagation in position space. The regions of localization change from one step to another one (dynamical Anderson localization) which is in contrast of what was reported for classical walk. In classical walk, the localization is over and around the place where the walker starts its walk. But for large number of entangled qubits, its not fixed over a specific region. Here, we should make an exception for two entangled qubits case where there is a localization similar to classical walk. The dependency of symmetry of distribution on number of entangled qubits enabled us to recognize the contribution of coin-shift operator on this property. The deterministic behavior in position space for the walker was also another consequences of increment in number of the entangled qubits. It was highlighted that based on considered criteria in this paper, walks with three entangled qubits have better efficiency comparing to other cases. Here, we should point it out that walks with larger number entangled qubits offer more diverse programmability. This is coming from the fact that one can consider different coins acting on each entangled qubits. This could change the distribution in coin space considerably resulting into different evolution in position space as well.

\section{Appendix: Analytical description}

In this section, we give an analytical description for the evolution of walk with different number of entangled qubits. The spatial Discrete Fourier Transform of $\Psi_{T}(x)$ is given by  

\begin{eqnarray}
	\Psi_{T}(k) = \sum_{x\in \mathbb{Z}}\Psi_{T}(x)e^{ikx}.
\end{eqnarray}

Since the walks start at the origin of the position space, we obtain 

\begin{eqnarray}
	\Psi_{0}(k) = \Psi_{0}(0).  
\end{eqnarray}

The evolution of walk will be then 
\begin{eqnarray}
	\Psi_{T}(k) =U(k)^{T} \Psi_{0}(k),  
\end{eqnarray}
in which 
\begin{eqnarray*}
	U(k) & = & U_{s}(\frac{k}{2})^ {\otimes n}=
	\bigg(\begin{bmatrix}e^{\frac{i k}{2}} & 0 \\0 & e^{-\frac{i k}{2}}\end{bmatrix} \widehat{C'} \bigg)^ {\otimes n}.  
\end{eqnarray*}

It is straight forward to find eigenvalues of $ U_{s}(k/2)$ as 

\begin{eqnarray*}
	\lambda_{\pm} & = & \pm \sqrt{1-\frac{ \sin^{2} (\frac{k}{2})}{2}}+i \frac{\sqrt{2}}{2} \sin^{2} (\frac{k}{2})=\pm e^{\pm\frac{i \phi(k)}{2}}.  
\end{eqnarray*}
where $\phi(k)=2 \sin^{-1} (\frac{\sin (k/2)}{\sqrt{2}})$. Since $U(k)=U_{s}(k/2)^{\otimes n}$, one can obtain the eigenvalues of $U(k)$ as 

\begin{eqnarray}
	\left\{
	\begin{array}{c} 
		\Lambda_{1} = \lambda_{+}^n
		\\[0.1cm]
		\Lambda_{2} = \lambda_{+}^{n-1} \lambda_{-}
		\\
		.
		\\
		.
		\\
		.
		\\
		\Lambda_{2^n-1} = \lambda_{-}^{n-1} \lambda_{+}
		\\[0.1cm]
		\Lambda_{2^n}= \lambda_{-}^n 
	\end{array}    \label{E1}   
	\right.  ,                  
\end{eqnarray}
and the eigenvectors are given by 

\begin{eqnarray}
	\left\{
	\begin{array}{c} 
		V_{1}(k)= v_{+}^{\otimes n}
		\\ [0.1cm]
		V_{2}(k)= v_{+}^{\otimes n-1} \otimes v_{-}
		\\
		.
		\\
		.
		\\
		V_{2^n-1}(k)= v_{-}^{\otimes n-1} \otimes v_{+}  
		\\
		V_{2^n}(k)= v_{-}^{\otimes n}    
	\end{array}  
	\right.  ,      \label{EV1}            
\end{eqnarray}
where $v_{\pm}$ are eigenvectors of $U_{s}(k/2)$ obtained as 

\begin{eqnarray*}
	v_{\pm} & = & \frac{1}{\sqrt{N_{\pm}}} \begin{bmatrix}e^{\frac{i k}{2}},\gamma_{\pm} & \end{bmatrix}  
\end{eqnarray*}
in which 

\begin{eqnarray*}
	N_{\pm} & = & 2-2\gamma_{\pm} \cos \frac{k}{2}  
	\\
	\gamma_{\pm} & = & -\cos \frac{k}{2} \pm \sqrt{1+ \cos^2\frac{k}{2}}  
\end{eqnarray*}

Given eigenvalues and eigenvectors, one can write $\Psi_{T}(k)$ as 

\begin{eqnarray*}
	\Psi_{T}(k) = U(k)^{T} \Psi_{0}(k) = \sum_{z=1}^{2^n} \Lambda_{z}^{T} \langle V_{z}(k), \Psi_{0}(k)\rangle V_{z}(k).   
\end{eqnarray*} 

By inverse Fourier transformation, we can find the amplitude of the wave function of walker at the
position $x$ and the step $T$ as

\begin{eqnarray}
	\Psi_{T}(x)=\int_{0}^{2\pi} e^{-ikx} \Psi_{T}(k) \frac{dk}{2\pi}, \label{PSI}     
\end{eqnarray}
and the probability of finding walker at position $x$ at step $T$ is obtained by $p_{T}(x)=\|\Psi_{T}(x)\|^2$ where $\| \|$ denotes vector norm. 

Let $m$ and $q$ be the numbers of $\lambda_{+}$ and $\lambda_{-}$ in calculation of a given eigenvalue ($m+q=n$), respectively, then one can obtain the eigenvalues in the following form 

\begin{eqnarray}
	\Lambda_{j} & = & \left\{
	\begin{array}{ccc} 
		(-1)^{m} e^{\frac{m-q}{2} i \phi(k)} &  & \text{   \textit{for even n}}
		\\[0.4cm]
		(-1)^{m+1} e^{\frac{m-q}{2} i \phi(k)} &  & \text{   \textit{for odd n}}  
	\end{array}  
	\right.  .     \label{E2}              
\end{eqnarray}

It should be noted that for $n$ number of the qubits, there are $2^n$ eigenvalues and the possible forms are given by the above. Accordingly, the eigenvectors are also calculated as 

\begin{eqnarray}
	V_{j}(k) & = & \frac{1}{N_{+}^{\frac{m}{2}}N_{-}^{\frac{q}{2}}} \begin{bmatrix}e^{\frac{i n k}{2}},...,\gamma_{+}^{m}\gamma_{-}^{q} \end{bmatrix}.     \label{EV2}              
\end{eqnarray}

The considered initial state is

\begin{eqnarray}
\Psi_{0}(0) & = & \begin{bmatrix}\frac{\sqrt{2}}{2},0,...,0,\frac{\sqrt{2}}{2}\end{bmatrix}^t 
\end{eqnarray}
where $t$ denotes the transpose. Using this initial state with Eqs. \eqref{E2}-\eqref{EV2}, we obtain the integrand of \eqref{PSI} as 

\begin{widetext}
	\begin{eqnarray}
		\Psi_{j} & = &  \left\{
		\begin{array}{ccc} 
			\frac{\sqrt{2}}{2}(-1)^{T m}e^{-ikx} e^{T\frac{m-q}{2} i \phi(k)} \frac{e^{-\frac{i n k}{2}}+\gamma_{+}^{m}\gamma_{-}^{q}}{N_{+}^{\frac{m}{2}}N_{-}^{\frac{q}{2}}} \frac{V_{j}}{2\pi}   &  & \text{\textit{for even n}}
			\\[0.4cm]
			\frac{\sqrt{2}}{2}(-1)^{T(m+1)}e^{-ikx} e^{T\frac{m-q}{2} i \phi(k)} \frac{e^{-\frac{i n k}{2}}+\gamma_{+}^{m}\gamma_{-}^{q}}{N_{+}^{\frac{m}{2}}N_{-}^{\frac{q}{2}}} \frac{V_{j}}{2\pi} &  & \text{\textit{for odd n}}  
		\end{array}  
		\right.  .     \label{IntG}           
	\end{eqnarray}
\end{widetext}

Once again, we emphasize that there are $2^n$, $\Psi_{j}$ which their general forms are given by Eq. \eqref{IntG}.

For even number of entangled qubits with $m \neq q$, by using the method of stationary phase \cite{Nayak,Liu}, we can find 

\begin{eqnarray*}
	&   & \int_{0}^{2\pi} \Psi_{j} \frac{dk}{2\pi} \rightarrow O(t^{\frac{-1+s}{s}}) \text{ as } t\rightarrow \infty
\end{eqnarray*}
where $s$ is the highest order of the stationary points. Evidently, if the eigenvalues are independent of $k$, the corresponding $\Psi_{j}$ contributes significantly to the limiting amplitude. Therefore, for even number of entangled qubits, it seems that $m = q$ are dominant terms in the limiting amplitude. For $m = q$, we obtain

\begin{equation}
	\Psi_{h} = \frac{\sqrt{2}(-1)^{\frac{T n}{2}} e^{-ikx}}{4\pi} \left(\frac{e^{-\frac{i n k}{2}}+(-1)^{\frac{n}{2}}}{(N_{+}N_{-})^{\frac{n}{4}}}\right) V_{h}(k),   \label{psi1}   
\end{equation}
where $h$ denotes cases with $m=q$. The probability of finding walker for arbitrary position would then be given by 

\begin{eqnarray*}
	p_{T}(x,n)=\bigg\| \int_{0}^{2\pi} \sum_{h} \Psi_{h} dk \bigg\|^2,
\end{eqnarray*}
where the summation is over all the possible combinations in which $m=q$. The number of $\Psi_{h}$ could be obtained by $\frac{n!}{((\frac{n}{2})!)^2}$.

The obtained integrand in Eq. \eqref{psi1} has dependency of $\Psi_{h} \propto O(1/(N_{+}N_{-})^{\frac{n}{2}})$. Considering that $N_{+}N_{-}=6 + 2 \cos k > 1$, the amplitude of the probability density at the origin ($x=0$) decays very fast as number of entangled qubits increases. The only exception is for $n=2$ where this amplitude is significant resulting into a major peak at the origin. This is in agreement with our graphical presentations in Fig. \ref{Fig1}. Accordingly, one should look to other amplitudes of $\Psi_{j}$ for finding the behavior of the walker. To do so, we go back to method of stationary phases \cite{Nayak,Pathak}. We rewrite 

\begin{eqnarray}
	\Psi_{T}(x)=\int_{-\pi}^{-\pi} e^{i T \Phi(k)} g(k) \frac{dk}{2\pi}, \label{PSI1}     
\end{eqnarray}
where $\Phi(k)= \frac{m-q}{2} \phi(k)- \frac{x}{T}k$ and $g(k)$ is a smooth function given by 

\begin{eqnarray}
	g(k) & = &  \left\{
	\begin{array}{ccc} 
		\frac{\sqrt{2}}{2}(-1)^{T m} \frac{e^{-\frac{i n k}{2}}+\gamma_{+}^{m}\gamma_{-}^{q}}{N_{+}^{\frac{m}{2}}N_{-}^{\frac{q}{2}}} V_{j}   &  & \text{\textit{for even n}}
		\\[0.4cm]
		\frac{\sqrt{2}}{2}(-1)^{T(m+1)} \frac{e^{-\frac{i n k}{2}}+\gamma_{+}^{m}\gamma_{-}^{q}}{N_{+}^{\frac{m}{2}}N_{-}^{\frac{q}{2}}} V_{j} &  & \text{\textit{for odd n}}  
	\end{array}  
	\right.      \label{IntG1}           
\end{eqnarray}

In the limit of large $T$ and by using method of stationary phase, we can find 

\begin{eqnarray}
	\Psi_{T}(x) \approx \sqrt{\frac{2\pi}{T|\Phi''(k_{0})|}} \Re[g(k_{0})e^{i T \Phi(k_{0})-i \frac{\pi}{4}}], \label{PSI11}     
\end{eqnarray}
for $-\frac{1}{\sqrt{2}}< \frac{x}{T}<\frac{1}{\sqrt{2}}$ with prime denoting derivation with respect to $k$ and $k_{0}$ is the root of $\Phi'(k_{0})=0$. It is a matter of calculation to show 

\begin{eqnarray}
	\Phi'(k) & = & \frac{(m-q)\cos\frac{k}{2}}{2 \sqrt{2-\sin^2 \frac{k}{2}}}-\frac{x}{T}, \label{phi1} 
	\\[0.2cm]
	\Phi''(k) & = & -\frac{(m-q)\sin\frac{k}{2}}{\sqrt{2} (3+\cos k)^{\frac{3}{2}}},  \label{phi2}
	\\[0.2cm]
	k_{0} & = & \left\{
	\begin{array}{c} 
		- 2 \cos^{-1} \bigg(\pm \frac{2x}{T\sqrt{(m-q)^2-4(\frac{x}{T})^2}} \bigg)
		\\[0.4cm]
		2 \cos^{-1} \bigg(\pm \frac{2x}{T\sqrt{(m-q)^2-4(\frac{x}{T})^2}} \bigg)
	\end{array}  
	\right.  \label{k0}
\end{eqnarray}

It should be noted that here, $m \neq q$. According to method of the stationary phases \cite{Nayak}, since for large $T$, the integral in \eqref{PSI1} is rapidly oscillating, the contributions from adjacent subintervals nearly cancel each other out, and the major contribution comes from the region where the oscillations are least rapid. The regions of least rapid oscillations are marked by $\Phi'(k_{0})=0$ which are known as the stationary points. Therefore, significant contributions come from a small interval around the stationary points. In our calculations, there are four distinguishable $k_{0}$ available for our system of equations. This denotes the presence of four distinguishable points where rapid oscillations are the least and have the highest contributions to $\Psi_{T}(x)$, hence probability density. This is in agreement with our plotted diagrams in Fig. \ref{Fig1} where four distinguishable peaks are observed in probability density distribution. 

Dependency of the obtained $\Phi(k)$, $\Phi''(k)$ and $k_{0}$ on $m$ and $q$ show that the regions of least rapid oscillations are functions of the number of entangled qubits and the dominant terms are the ones where $m \rightarrow q$. This is because for the limit $m \rightarrow q$, $\Phi''(k_{0})$ and $\arrowvert \Phi(k_{0}) \arrowvert$ become minimum. The asymmetric probability density distribution that was observed for odd qubits (absent for even qubits) is due to $(-1)^{T(m+1)}$ factor given in Eq. \eqref{IntG1}. Careful examination of Eq. \eqref{IntG1} shows that eigenvalues where $m-q>0$ and ones with $m-q<0$ have opposite signs for odd number of qubits whereas, they have identical signs in case of even qubits. Therefore, we see a symmetric probability density distribution for even number of entangled qubits and aymmetric one for odd number of entangled qubits. To achieve symmetrical probability density distribution for odd number of entangled qubits, one should change the initial state which essentially means changing $\langle V_{z}(k), \Psi_{0}(k)\rangle$ in our calculations.

\newpage


\begin{thebibliography}{99}  
	
	
\bibitem{Childs}	
    A. ~M. ~Childs, Phys.\ Rev.\ Lett.\  \textbf{102}, 180501 (2009).	
	
 \bibitem{Lovett}
    N. ~B. ~Lovett et al, Phys.\ Rev.\ A\ \textbf{81}, 042330 (2010).	
    

\bibitem{Ambainis}     
   A. Ambainis, Int. J. Quantum Inf. \textbf{4}, 507 (2003). 
 
\bibitem{Shenvi}
   N. ~Shenvi et al, Phys.\ Rev.\ A\ \textbf{67}, 052307 (2003).   
 
	 
\bibitem{Mohseni}
   M.~Mohseni et al, J.\ Chem.\ Phys.\ \textbf{129}, 174106 (2008).  
	

\bibitem{Innocenti}     
   L. ~Innocenti et al, Phys.\ Rev.\ A\ \textbf{96}, 062326 (2017).
    
    
\bibitem{Paparo}     
   G. ~Paparo et al, Phys. Rev. X \textbf{4}, 031002 (2014).   
    

\bibitem{Schuld1}     
   M. ~Schuld, I. ~Sinayskiy, and F. ~Petruccione, Quant.\ Info.\ Process.\ \textbf{13}, 2567 (2014).

\bibitem{Schuld2} 
   M. ~Schuld, I. ~Sinayskiy, and F. ~Petruccione, Phys.\ Rev.\ A\ \textbf{89}, 032333 (2014).
   
   
\bibitem{Dadras}
  S. ~Dadras et al, Phys.\ Rev.\ Lett.\ \textbf{121}, 070402 (2018).

\bibitem{Barkhofen}  
  S. ~Barkhofen et al, Phys.\ Rev.\ Lett.\ \textit{121}, 260501 (2018). 


\bibitem{Schreiber}
  A. ~Schreiber et al, Phys.\ Rev.\ Lett.\ \textbf{104}, 050502 (2010).
  
\bibitem{Bian}
  Z. ~Bian et al, Phys.\ Rev.\ Lett.\ \textbf{114}, 203602 (2015).

\bibitem{Zahringer}
  F. ~Zahringer et al, Phys.\ Rev.\ Lett.\ \textbf{104}, 100503 (2010). 

\bibitem{Karski}
  M. ~Karski et al, Science\ \textbf{325}, 174 (2009).

   

\bibitem{Collins}    
   D. ~Collins et al, Phys.\ Rev.\ Lett.\ \textbf{88}, 040404 (2002).


\bibitem{Vertesi}
  T. ~V\'ertesi et al, Phys.\ Rev.\ Lett. \textbf{104}, 060401 (2010).

\bibitem{Martin} 
  A. Martin et al, Phys.\ Rev.\ Lett.\ \textbf{118}, 110501 (2017).
  

\bibitem{Cerf} 
  N. J. ~Cerf et al, Phys.\ Rev.\ Lett.\ \textbf{88}, 127902 (2002).    
   

\bibitem{Venegas}
  S. ~E. ~Venegas-Andraca et al, New\ J.\ Phys.\ \textbf{7}, 221 (2005).

\bibitem{Liu}    
  C. ~Liu and N. ~Petulante, Phys.\ Rev.\ A\ \textbf{79}, 032312 (2009).
  
\bibitem{Liu2012}
  C. ~Liu, Quant.\ Info.\ Proc.\ \textbf{11}, 1193 (2012). 
  
\bibitem{Panahiyan} 
  S. ~Panahiyan and S. Fritzsche, [arXiv: 1810.11020].    


\bibitem{Sanders} 
  B. ~C. ~Sanders et al, Phys.\ Rev.\ A \textbf{67}, 042305 (2003).

\bibitem{Flurin} 
  E. ~Flurin et al, Phys.\ Rev.\ X \textbf{7}, 031023 (2017).


\bibitem{Casabone} 
  B. ~Casabone et al, Phys.\ Rev.\ Lett.\ \textbf{114}, 023602 (2015).  

  
\bibitem{Brun}  
  T. ~A. ~Brun, H. ~A. ~Carteret, and A. ~Ambainis, Phys.\ Rev.\ Lett.\ \textbf{91}, 130602 (2003).
   

\bibitem{Portugal}    
  R. ~Portugal, \textit{Quantum Walks and Search Algorithms}, Springer, New York (2013).   
   

\bibitem{Keating}  
  J. P. Keating et al, Phys. Rev. A \textbf{76}, 012315 (2007). 

   
\bibitem{Nayak}    
  A. ~Nayak and A. ~Vishwanath, e-print arXiv:quant-ph/0010117.  

\bibitem{Pathak}   
  P. K. Pathak and G. S. Agarwal, Phys.\ Rev.\ A\ \textbf{75}, 032351 (2007).    
 
    
\end{thebibliography}
\end{document}